\documentclass[a4paper, aps, prb, reprint, superscriptaddress]{revtex4-1}
\usepackage{graphicx}
\usepackage{floatrow}

\newfloatcommand{capbtabbox}{table}[][\FBwidth]
\usepackage[caption=false,font=footnotesize]{subfig}

\usepackage{epstopdf}
\usepackage{amsmath}
\usepackage{amssymb}
\usepackage{color}
\usepackage{multirow}
\newcommand{\Rsq}{R_\Box}
\newcommand{\Tc}{T_\mathrm{c}}

\newcommand{\unit}[1]{\ensuremath{\, \mathrm{#1}}}
\newcommand{\unitT}[1]{\ensuremath{\, \textrm{#1}}}

\begin{document}
\title{Electrodynamic response and local tunnelling spectroscopy of strongly disordered superconducting TiN films}
\author{P. C. J. J. Coumou}\email{p.c.j.j.coumou@tudelft.nl}
\affiliation{Kavli Institute of Nanoscience, Delft University of Technology, Delft, The Netherlands}
\author{E. F. C. Driessen}\email{eduard.driessen@cea.fr}
\affiliation{SPSMS/LaTEQS, UMR-E 9001, CEA-INAC and Universit\'e Joseph Fourier, Grenoble, France}

\author{J. Bueno}
\affiliation{SRON National Institute for Space Research, Utrecht, The Netherlands}

\author{C. Chapelier}
\affiliation{SPSMS/LaTEQS, UMR-E 9001, CEA-INAC and Universit\'e Joseph Fourier, Grenoble, France}
\author{T. M. Klapwijk}
\affiliation{Kavli Institute of Nanoscience, Delft University of Technology, Delft, The Netherlands}
\affiliation{Physics Department, Moscow State Pedagogical University, 119991 Moscow, Russia}
\date{\today}

\begin{abstract}
We have studied the electrodynamic response of strongly disordered superconducting TiN films using microwave resonators, where the disordered superconductor is the resonating element in a high-quality superconducting environment of NbTiN. We describe the response assuming an effective pair-breaking mechanism modifying the density of states, and compare this to local tunnelling spectra obtained using scanning tunnelling spectroscopy. For the least disordered film ($k_\mathrm{F}l=8.7$, $R_\mathrm{s}=13~\Omega$), we find good agreement, whereas for the most disordered film ($k_\mathrm{F}l=0.82$, $R_\mathrm{s}=4.3~\mathrm{k}\Omega$), there is a strong discrepancy, which signals the breakdown of a model based on uniform properties.
\end{abstract}

\maketitle

The evolution of the superconducting state by increasing disorder is an important and complex problem. Many processes are present simultaneously, and the predominant experimental outcome is that the system evolves towards an insulating state\cite{Gantmakher:2010il} (\emph{i.e.} increasing resistance with decreasing temperature), while at the same time maintaining properties reminiscent of Cooper-pairs\cite{Sacepe:2011jm}. Recently, we have carried out\cite{Driessen:2012gx} a systematic study of the microwave-response of TiN films with values of $k_\mathrm{F} l$ slightly above $1$. These results have been compared with a recent theory of Feigel'man and Skvortsov,\cite{Feigelman:2012fp} designed to describe the effect of disorder and the emergence of mesoscopic inhomogeneities on the quasiparticle density of states in the superconducting state. It was found that this theory strongly underestimates the effects of disorder observed in our experiments. Instead we have used a heuristic model which treats the effect of disorder as a pair-breaking parameter that is not a priori related to the level of disorder. Here, we extend the experimental results to films with sheet resistance $R_\mathrm{s}$ up to $4.3~\mathrm{k}\Omega$ and $k_\mathrm{F} l$-values down to $0.82$. We observe a similar trend as previously reported, with a strong increase of the effective pair-breaking parameter for decreasing $k_\mathrm{F} l$. Since the analysis assumes a modified density-of-states we supplement our analysis with data derived from local tunnelling spectroscopy on the same films. For the least disordered film ($k_\mathrm{F}l=8.6$, $R_\mathrm{s}=13~\Omega$), we find good agreement, but for the most disordered film ($k_\mathrm{F}l = 0.82$, $R_\mathrm{s}=4.3~\mathrm{k}\Omega$), there is a strong discrepancy between the locally measured density of states and the density of states that has been assumed to describe the electrodynamic response.

To measure the electrodynamics of superconductors, commonly coplanar waveguide (CPW) resonators are used, which are capacitively coupled to a readout transmission line, where  the resonator, the ground plane, and the transmission line are all made of the same material (i.e. the superconductor under study). In the regime where $k_\mathrm{F}l\approx 1$, this approach becomes unfeasible. The large sheet resistance in this regime ($\Rsq > 1\unit{k\Omega}$) translates into an equally large sheet inductance $L_\mathrm{S}\approx\hbar\Rsq/\pi\Delta$, with $\Delta$ the superconducting pairing potential. For increasing disorder, it therefore becomes increasingly more difficult to match the characteristic impedance of the transmission line $Z\approx\sqrt{L/C}$ to the $50\unit{\Omega}$ impedance of the microwave equipment, with $L$ and $C$ the total inductance and capacitance, respectively. This results in standing waves in the transmission line and spurious resonances.

\begin{figure}[tb]
\centering
\includegraphics[width=\linewidth]{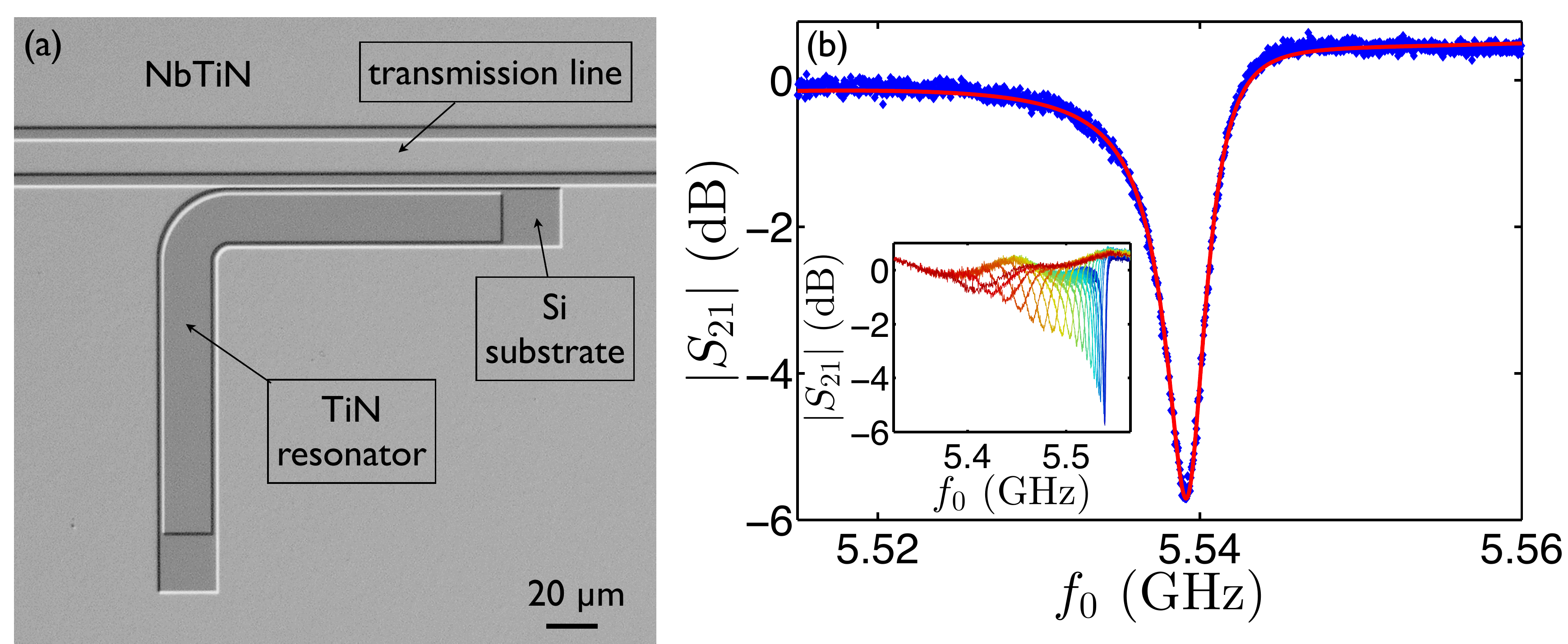}
\caption{(Color online) (a) Optical microscope image of the resonator design used in this paper. A TiN half-wave resonator is embedded in a NbTiN ground plane, and capacitively coupled to the NbTiN transmission line. (b) Transmission line throughput $S_\mathrm{21}$ as a function of microwave frequency, for a resonator from film E (blue dots). The solid red curve is a fit of Eq.~(\ref{eq:S21}) to the measured data, used to extract the resonance frequency $f_0$. The design allows to follow the resonance curve for a temperature range going from $T=50\unit{mK}$ (blue) to $T=417\unit{mK}$ (red) (inset).}
\label{fig:1}
\end{figure}

To address this issue, we have adapted a resonator design by Diener \emph{et al.}\cite{Diener:2012hk} In our design, a half-wave CPW resonator of TiN is embedded in an environment of 100\unit{nm} thick NbTiN. First, the resonator structure is defined in the TiN film, using e-beam lithography and reactive ion etching. In a second step, the NbTiN layer is deposited on a lift-off mask, using DC magnetron sputtering, at a DC power of 300~W, using an argon flow of 100~sccm and a nitrogen flow of 4~sccm, with a chamber pressure of 4~mTorr.\cite{Iosad:2000gz} Lift-off of the NbTiN is done in acetone.\footnote{In the case of film C, the fabrication process was slightly different, resulting in the TiN layer  still being present beneath the NbTiN layer.} Figure~\ref{fig:1}(a) shows an optical micrograph of one of the resonators.

Stronger disorder also results in a lower internal quality factor $Q_\mathrm{i}$ of the resonator.\cite{Coumou:gg} Changes in the properties of the resonators are best assessed when $Q_\mathrm{i}\approx |Q_\mathrm{c}|$, with $Q_\mathrm{c}$ the complex quality factor corresponding to the (dominantly capacitive) coupling to the transmission line.\cite{Zmuidzinas:2012wj} Maximal coupling (minimal $|Q_\mathrm{c}|$) is obtained by positioning the resonators such that half of their length is adjacent to the transmission line. This design yields a coupling quality factor of $|Q_\mathrm{c}| \approx 3\cdot 10^3$, which should be compared to a typical $Q_\mathrm{i} \le 10^4$. Detailed dimensions of the sample design can be found in the supplementary information.\cite{supplementary}
Figure~\ref{fig:1}(b) shows a typical resonance curve of a resonator made from film E using this design. The resonance dip is clearly visible at the lowest temperature of 50~mK, and can be followed up to a temperature of $417\unit{mK}$ (inset).

%
%
%
%


\begin{table}[!ht]
\renewcommand{\arraystretch}{1}
\caption{Parameters of the films. The values of $\Delta_0$ and $\alpha$ are extracted from the fits to the measurements of the electrodynamics.}
\label{Table:1}
\centering
\begin{tabular}{c c c c c c c c}
\hline\hline
Film & Thickness& $\Tc$ & $R_s$ & $l$ & $k_Fl$ & $\Delta_0$ & $\alpha/\Delta_0$ \\
 & $(\unit{nm})$ & $(\unit{K})$ & $(\unit{k\Omega})$ & $(\unit{\textrm{\AA}})$ & & $(\mu\mathrm{eV})$ & \\
\hline
C & $4.0$ & $0.70$ & $4.3$ & $0.96$ & $0.82$ & $307$ & $0.65$\\
E & $4.5$ & $0.99$ & $3.0$ & $1.2$ & $1.1$ & $277$ & $0.49$ \\
G & $5.0$ & $1.5$ & $1.5$ & $1.9$ & $1.8$ & $288$ & $0.23$ \\
\hline
I & $89$ & $3.5$ & $0.013$ & $7.3$ & $8.6$ & $613$ & $0.11$ \\
\hline \hline
\end{tabular}
\end{table}

The TiN films used in this study are grown using plasma-assisted atomic-layer deposition (ALD). The ALD technique is based on sequential self-terminating gas-solid reactions at the film surface, using two gaseous precursors. We make use of the precursors TiCl$_4$ and a plasma of H$_2$ and N$_2$, which react into TiN and gaseous HCl. The films are deposited on highly resistive ($\rho > 10\unit{k\Omega cm}$) Si(100) substrates that have a thin surface layer of native silicon oxide. 
The level of disorder is tuned by changing the thickness $d$ of the films\cite{Driessen:2012gx,Coumou:gg} (see Table~\ref{Table:1}, film C, E, and G). The resulting films are not directly comparable to the previously studied films (notably the critical temperature is slightly higher), due to a maintenance update of the machine. The thick film I is part of the previous deposition series.\footnote{This film is labelled $E$ in Ref.\onlinecite{Driessen:2012gx}}.

Our ALD process yields thin films with a typical RMS surface roughness of $2\unitT{\AA}$, determined by atomic force microscopy. Cross-sectional transmission electron microscopy of film C reveals that the TiN layer is poly-crystalline with a typical grain size of $5\unit{nm}$. The thickness of the various films is estimated using an average growth rate of $0.5\unitT{\AA}/\mathrm{cycle}$.\cite{supplementary}


Hall bar structures patterned in each film are used to measure the sheet resistance as a function of temperature $T$, and the Hall voltage $V_\mathrm{H}(B)$ as a function of magnetic field. The critical temperature $\Tc$ is determined from the the resistive transition.\cite{supplementary} For our films, the superconductor-to-insulator transition (SIT) occurs at a sheet resistance between 4.3 and $16\unit{k\Omega}$, consistent with previously reported values for the SIT in TiN.\cite{Baturina:2008ww} The sheet resistance, elastic scattering length $l$, and $k_\mathrm{F}l$ are determined at a temperature of $10\unit{K}$, using free-electron theory.\cite{supplementary} The Ioffe-Regel parameter $k_\mathrm{F}l$ decreases monotonously with film thickness, to values $<1$ for film C, indicating strong electron localization. The films C-G are part of a larger series, described in the supplementary information.\cite{supplementary}

\begin{figure}[!ht]
\centering
\includegraphics[width=\linewidth]{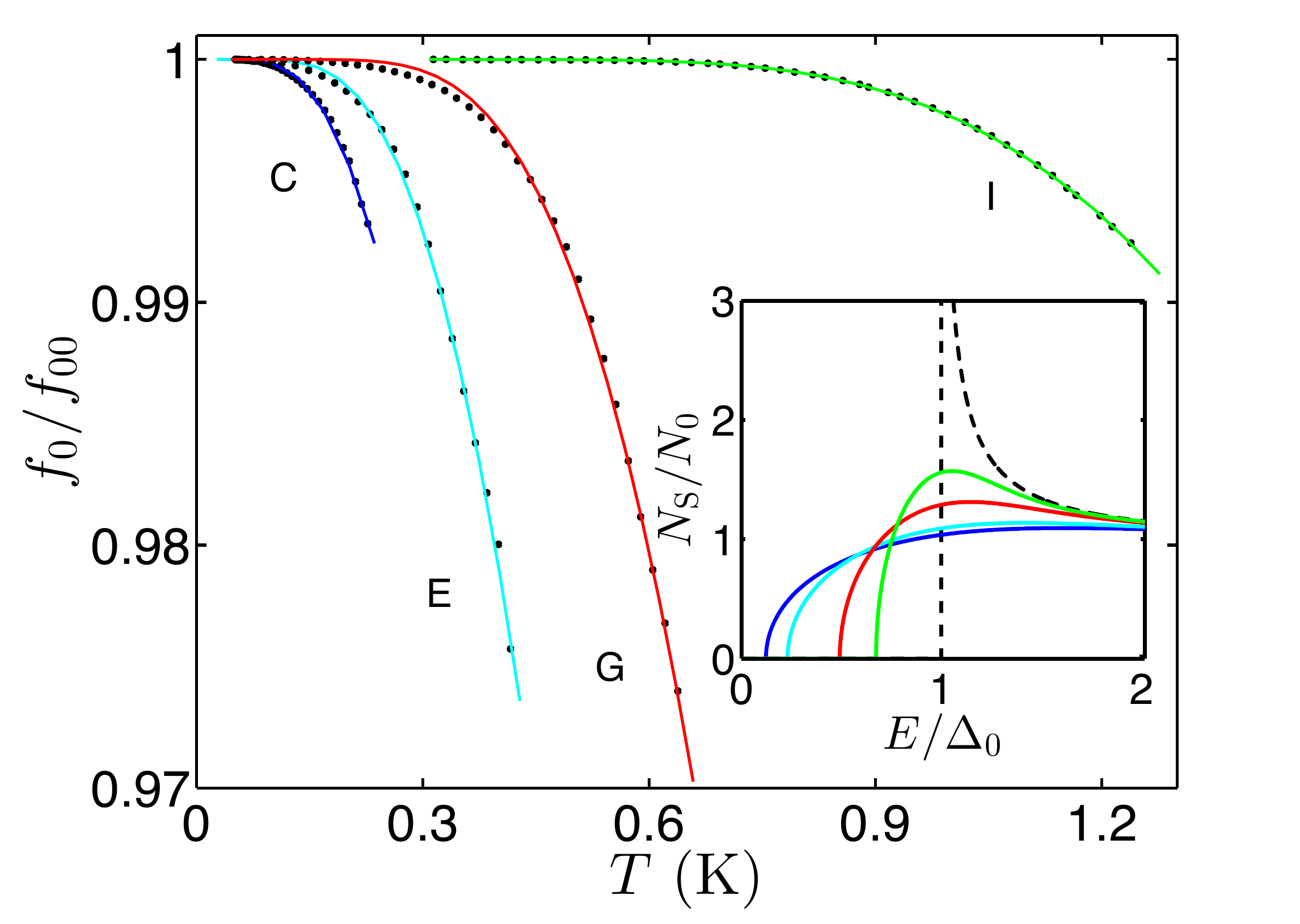}
\caption{(Color online) Measured resonance frequency as a function of temperature for films C, E, G, and I. The solid curves are a fit of the model to the data (see text for details). The inset shows the superconducting densities of states corresponding to these fits, compared to the BCS density of states (dashed curve) as a comparison.}
\label{fig:2}
\end{figure}

We have fabricated resonator samples from these films, using the design described before. The samples are wire bonded to coaxial connectors of a cryogenic microwave set-up, and cooled to a base temperature of $50\unit{mK}$ using a pulse-tube pre-cooled adiabatic-demagnetization refrigerator. A microwave signal from a vector network analyzer is fed to the sample through coaxial cables. The amplified forward power transmission $S_{21}$ of the transmission line is recorded as a function of temperature and microwave frequency (Fig.~\ref{fig:1}(b)).  A small linear background is observed in these measurements, due to small imperfections in the calibration of the setup. We subtract this background from the data, and extract the resonance frequency $f_0$ by fitting the relation\cite{Khalil:2012jr}
\begin{equation}
\frac{1}{S_{21}} = 1+\frac{Q_\mathrm{i}/Q_\mathrm{c}}{1+2iQ_\mathrm{i}\frac{f-f_0}{f_0}},
\label{eq:S21}
\end{equation}
where $f$ is the microwave frequency. The solid curve in Fig.~\ref{fig:1}(b) is a fit of this line shape to the measured curve. Different resonators on the same chip show an identical temperature dependence. For increasing temperature, the resonance frequency $f_0$ is decreasing, reflecting a change in the sheet inductance $L_\mathrm{S}$ of the superconductor. Figure~\ref{fig:2} shows the change in resonance frequency as a function of temperature for all four measured films.

 We model the microwave response using a description of the superconducting state, in which the superconductor is homogeneously weakened by a disorder-dependent pair breaking parameter $\alpha$, similar to the effect of magnetic impurities.\cite{Driessen:2012gx} We assume homogeneous superconductivity, and describe the superconducting state using the Usadel equation
\begin{equation}
iE\sin\theta + \Delta\cos\theta - \alpha\sin\theta\cos\theta = 0,\label{eq:usadel}
\end{equation}
where $E$ is the quasiparticle energy,  $\sin\theta$ and $\cos\theta$ are the quasi-classical Green's functions, and $\Delta$ is the pairing amplitude that is determined self-consistently for each temperature and value of $\alpha$.\cite{supplementary} The effect of the pair-breaking parameter is to smoothen the coherence peak in the quasiparticle density of states, as shown in the inset of Fig.~\ref{fig:2}.

In contrast to our earlier work, here we take $\Delta_0$, the pairing amplitude at zero temperature, as a free parameter in the model. This choice is motivated by the observation that there is no a priori relationship between $T_\mathrm{c}$ and $\Delta_0$ for this kind of films.\cite{Sacepe:2008jx}

After solving the Usadel equation (\ref{eq:usadel}), the complex conductivity $\sigma = \sigma_1-i\sigma_2$ is determined using a generalized form of the Mattis-Bardeen equations, given by Nam.\cite{Nam:1967ue} The resonance frequency $f_0$ is only dependent on $\sigma_2$,\cite{supplementary}
\begin{equation}
f_0 \propto \frac{1}{\sqrt{L_\mathrm{S}(\sigma_2)}}.
\end{equation}
The solid curves in Fig.~\ref{fig:2} show the best fits of this model to the measured resonance frequencies. To obtain these fits, we use the resonance frequency at the lowest measured temperature, $f_{00}$, as a scaling parameter. For a number of values of $\alpha$, we determine a least-squares fit using $\Delta_0$ as the fitting parameter. Finally, we determine the value of $\alpha$ where the residual error of the fit was minimal. The resulting values of $\alpha$ and $\Delta_0$ are presented in Table~\ref{Table:1}. As before, the trend that $\alpha$ increases with increasing disorder,\cite{Driessen:2012gx} is also observed in these films. The inset of Fig.~\ref{fig:2} shows the quasiparticle density of states $N_\mathrm{S}\propto \mathrm{Re}\ \cos\theta$, that corresponds to the fitted value of $\alpha$ for all four films.
Our model implies distinct predictions for the quasiparticle density of states of the film. To test these predictions, we independently probed the local density of states using scanning tunnelling spectroscopy (STS) on unpatterned pieces of films I and C.
 A piece of film I is mounted in a pumped-$^4$He bath cryostat and cooled down to a bath temperature of 1.35~K. Film C is mounted in an inverted $^3$He-$^4$He dilution refrigerator, and cooled down to a base temperature of 50~mK. A sharp Pt-Ir tip is brought in close proximity to the films, such that a tunnelling current of 1~nA is reached in both cases, at the maximum bias voltage of 2~mV for film I, and 500~$\mu$V for film C. A low-frequency modulation with an amplitude of 10~$\mu$V and 20~$\mu$V, respectively, is applied to the DC voltage bias, to measure the differential conductance using lock-in techniques. Tunnelling spectra are obtained by ramping the DC bias voltage, while keeping the distance between the tip and the sample fixed.

\begin{figure}[tb]
\centering
\includegraphics[width=\linewidth]{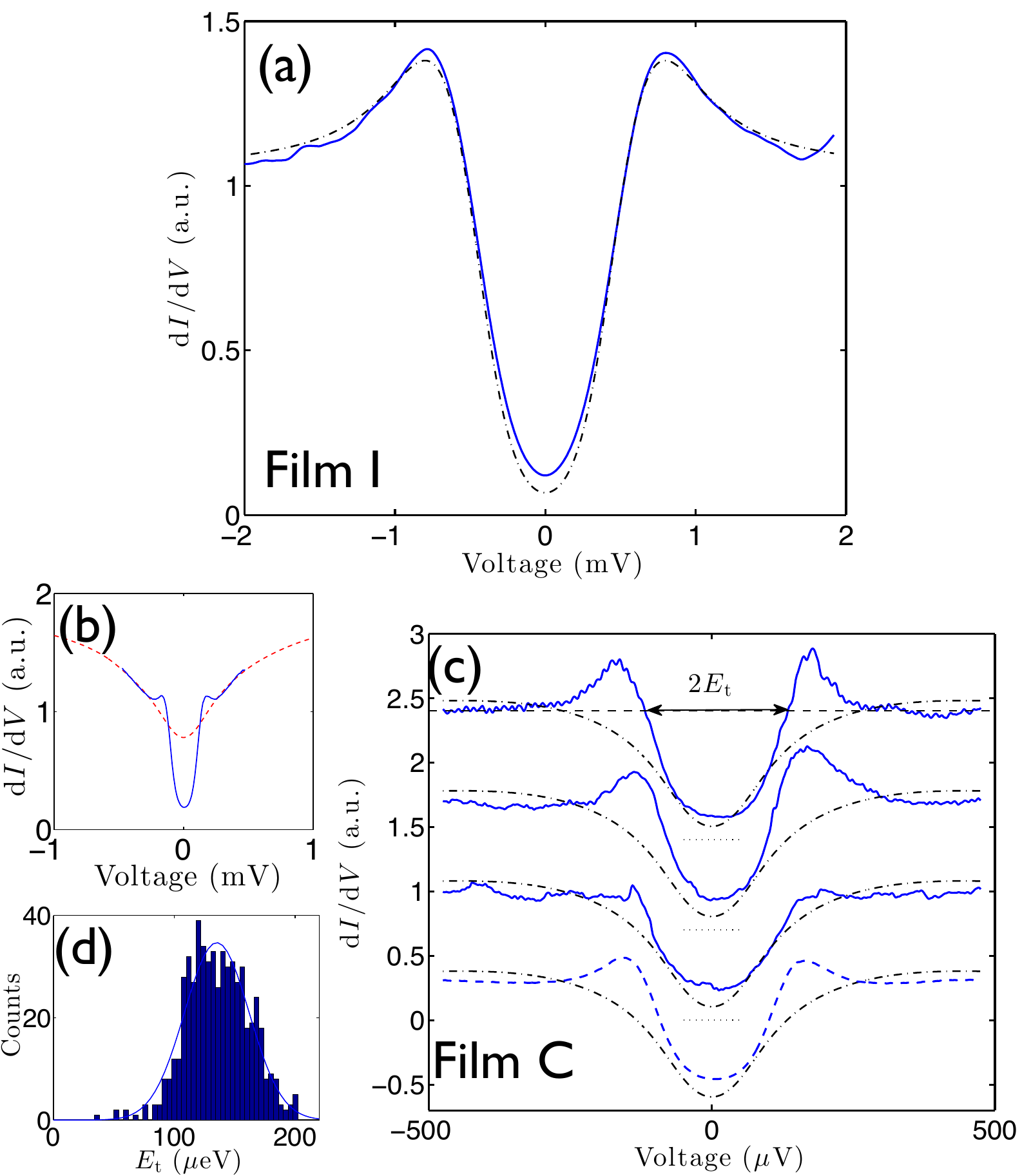}
\caption{(Color online) (a) Measured tunnelling spectrum of film I (solid curve). (b) Average tunnelling spectrum for film C (solid curve) at $T=50\unit{mK}$, together with the normal-state tunnelling spectrum at $T=1.4\unit{K}$ (red dashed curve). (c) Tunnelling spectra of film C at three different positions (solid curves), and the average tunnelling spectrum (dashed curve). The curves are shifted for clarity (the dotted lines give the zero-conductance levels). The normal-state background is removed from these measurements (see text for details). The dash-dotted black curves in (a) and (c) are the tunnelling spectra that are expected from the electrodynamics measurements. (d) Histogram of the characteristic energy scale $E_\mathrm{t}$ in the local spectra in film C. The gaussian fit (solid curve) is used to extract the mean value.}
\label{fig:sts1}
\end{figure}

Figure~\ref{fig:sts1}(a) shows the tunnelling spectrum on a single point of film I. The curve is an average of 50 consecutive spectra that were recorded at the same position.  The spectrum shows the characteristic BCS coherence peaks at a voltage of $800\unit{\mu V}$. A detailed scan over an area of 500$\times$500~nm$^2$ shows that the variation of the local density of states is within the noise level of the measurement for this film.

A different picture is obtained from the measurements on film C. On this film, spectra were obtained on a square grid of 26$\times$26 points in an area of 100$\times$100~nm$^2$. 
Figure~\ref{fig:sts1}(b) shows the average of the tunnelling spectrum measured at all positions (solid curve). Clearly, the BCS coherence peaks are visible, although they are not as pronounced as in film I. Besides that, the tunnelling conductivity is rising for increasing voltage. This rise is also observed at temperatures above $T_\mathrm{c}$, and is generally attributed to a depression of the normal-state tunnelling density of states at the Fermi level, caused by increased electron-electron interactions in the film.\cite{Altshuler:1985tw,Pierre:2001fo}. The red dashed curve in Fig.~\ref{fig:sts1}(b) shows the average background, measured over an area of 100$\times$100~nm$^2$ at a temperature of $T=1.4\unit{K}$. The depression of the tunnelling density of states is due to the energy cost of adding one electron to the film. Therefore, it is typical for tunnelling and we do not expect it to play a role in the electrodynamic response of the film. Consequently, we have divided all measured spectra by the average normal state tunnelling spectrum (red curve). 

Figure~\ref{fig:sts1}(c) shows the resulting tunnelling spectra, obtained at three different positions, where a small linear background, due to thermal drift in the tip-sample distance, is subtracted. From these curves, it is clear that the local properties of the film are varying on the 100~nm length scale. First of all, the height of the coherence peak is varying from point to point. There are locations where a distinct coherence peak is seen (top curves), as well as locations, where the coherence peak is absent. Besides that, the typical energy scale $E_\mathrm{t}$, defined as the crossing point with the normal-state tunnelling spectrum, is varying from point to point. On average, $E_\mathrm{t} = 134\unit{\mu eV}$, with a variation of $20\%$ (shown in Fig.~\ref{fig:sts1}(d)). This variation in energy scale and coherence peak height is of similar size as was observed in experiments in strongly disordered TiN,\cite{Sacepe:2008jx} InO$_x$,\cite{Sacepe:2011jm} and NbN.\cite{Lemarie:2013bk} It is consistent with numerical work relating it to homogeneous on-site disorder in competition with superconductivity.\cite{Ghosal:1998up,Ghosal:2002cn}

%

Let us now compare the local spectroscopy measurements to the results obtained from the  electrodynamics. The dash-dotted curves in Fig.~\ref{fig:sts1} show the  tunnelling spectra that were calculated using the values inferred from the electrodynamic response as shown in Table~\ref{Table:1}.  The curves are a convolution of the quasiparticle density of states calculated using Eq.~(\ref{eq:usadel}) with the derivative of the Fermi-Dirac distribution function describing thermal excitations in the normal-metal tip.
For the calculation of the curves, there is no fitting parameter, except for a scaling parameter which is proportional to the tunnelling conductance of the contact. The effective temperature of the tip is taken to be equal to the bath temperature ($T=1.35$~K) in case of film I, and assumed to be 300~mK in the case of film C, based on previous experience with this measurement setup.

In case of the least-disordered film I, there is a clear correspondence between the measurements and the predictions from the model: the size and width of the coherence peaks and their distance are accurately reproduced. In case of the most-disordered superconducting film C, the picture is different. The measured curves are not predicted by the model. First, in most of the spectra, coherence peaks are clearly visible, while they do not follow from the Usadel equation (\ref{eq:usadel}) for the parameters listed in Table~\ref{Table:1}. Secondly, the calculated curve is much smoother: even on positions where the coherence peaks are absent (bottom curve in Fig.~\ref{fig:sts1}(c)), there is still a clear onset of a gap, that is not reproduced by the model.

These differences may be attributed to the tunnelling measurement method. It was recently suggested,\cite{Sherman:2013ta} that the metallic tip is not only a passive probe of the tunnelling density of states, but may also influence the electron-electron interaction processes in the highly disordered film themselves. Although we can not rule out such an influence on a quantitative level, we believe that this cannot explain the observed spatial variation in the local superconducting density of states. 

In our view, the main result of the STS measurements are the local variations of the superconducting state.  In contrast, we apply for the electrodynamic properties a model, which assumes uniform properties, with an effective pair-breaking parameter. Since the wavelength of the microwaves used in our experiment is large enough to probe an average over the local landscape, an effective uniform model might still be appropriate. However, there is no a priori reason, why this is represented by an effective pair-breaker. In fact, the averaged tunnel spectrum, shown in Fig.~\ref{fig:sts1}(c) reveals a stronger superconducting state than is predicted by the electrodynamics measurements. A proper theoretical description of the microwave response will need to start with an inhomogeneous superconducting state. We are not aware of such a theory.  

Finally, in analysing the density of states as observed in our STS measurements of film C, we appear to observe an appreciable amount of subgap states. Also in this case we are reluctant to apply a model, which assumes a uniform superconductor with subgap states. For a non-uniform superconducting state the lateral proximity-effect will contribute to the locally measured density of states, which may appear as subgap states.\cite{Gueron:1996hw} Moreover, the influence of the electron-electron interactions, leading to the normal-state tunnelling density of states, on the superconducting state are not clear. A thorough understanding of the local tunnelling spectra in an inhomogeneous superconducting medium is needed, before one can infer a macroscopic quantity such as the response to microwaves. 

In summary, we have clearly shown that the microwave electrodynamics of strongly disordered superconductors into a regime up to $k_\mathrm{F}l  = 0.82$, can be measured by using a hybrid sample design. We have also shown that even in this regime we can describe the electrodynamic response with a model, which contains a disorder-dependent effective pair breaker. For the least disordered film ($k_\mathrm{F}l  = 8.6$), these results are consistent with the local tunnelling spectra.  For the most disordered film  however, a discrepancy is found, which signals the breakdown of applicability of a model based on averaged properties, due to the emergent electronic inhomogeneity.

We thank Misha Skvortsov, Benjamin Sac\'ep\'e,  Yuli Nazarov, and Jochem Baselmans for useful discussions, Frans Tichelaar for performing the TEM measurements, and Akira Endo and David Thoen for assistance in resonator design and process development. This research was funded by the Dutch Foundation for Research of Matter (FOM), and by the Agence Nationale de la Recherche under the programs POSTIT and QuDEC. This work has been supported as part of a collaborative
project, SPACEKIDS, funded via grant 313320 provided by the European
Commission under Theme SPA.2012.2.2-01 of Framework Programme 7. T.M.K. acknowledges financial support from the Ministry of Science and Education of Russia under contract No. 14.B25.31.0007. E.F.C.D. was financially supported by the CEA-Eurotalents program.

\bibliographystyle{apsrev4-1}

\begin{thebibliography}{22}%
\makeatletter
\providecommand \@ifxundefined [1]{%
 \@ifx{#1\undefined}
}%
\providecommand \@ifnum [1]{%
 \ifnum #1\expandafter \@firstoftwo
 \else \expandafter \@secondoftwo
 \fi
}%
\providecommand \@ifx [1]{%
 \ifx #1\expandafter \@firstoftwo
 \else \expandafter \@secondoftwo
 \fi
}%
\providecommand \natexlab [1]{#1}%
\providecommand \enquote  [1]{``#1''}%
\providecommand \bibnamefont  [1]{#1}%
\providecommand \bibfnamefont [1]{#1}%
\providecommand \citenamefont [1]{#1}%
\providecommand \href@noop [0]{\@secondoftwo}%
\providecommand \href [0]{\begingroup \@sanitize@url \@href}%
\providecommand \@href[1]{\@@startlink{#1}\@@href}%
\providecommand \@@href[1]{\endgroup#1\@@endlink}%
\providecommand \@sanitize@url [0]{\catcode `\\12\catcode `\$12\catcode
  `\&12\catcode `\#12\catcode `\^12\catcode `\_12\catcode `\%12\relax}%
\providecommand \@@startlink[1]{}%
\providecommand \@@endlink[0]{}%
\providecommand \url  [0]{\begingroup\@sanitize@url \@url }%
\providecommand \@url [1]{\endgroup\@href {#1}{\urlprefix }}%
\providecommand \urlprefix  [0]{URL }%
\providecommand \Eprint [0]{\href }%
\providecommand \doibase [0]{http://dx.doi.org/}%
\providecommand \selectlanguage [0]{\@gobble}%
\providecommand \bibinfo  [0]{\@secondoftwo}%
\providecommand \bibfield  [0]{\@secondoftwo}%
\providecommand \translation [1]{[#1]}%
\providecommand \BibitemOpen [0]{}%
\providecommand \bibitemStop [0]{}%
\providecommand \bibitemNoStop [0]{.\EOS\space}%
\providecommand \EOS [0]{\spacefactor3000\relax}%
\providecommand \BibitemShut  [1]{\csname bibitem#1\endcsname}%
\let\auto@bib@innerbib\@empty
\bibitem [{\citenamefont {Gantmakher}\ and\ \citenamefont
  {Dolgopolov}(2010)}]{Gantmakher:2010il}%
  \BibitemOpen
  \bibfield  {author} {\bibinfo {author} {\bibfnamefont {V.~F.}\ \bibnamefont
  {Gantmakher}}\ and\ \bibinfo {author} {\bibfnamefont {V.~T.}\ \bibnamefont
  {Dolgopolov}},\ }\href {\doibase 10.3367/UFNe.0180.201001a.0003} {\bibfield
  {journal} {\bibinfo  {journal} {Phys-Usp+}\ }\textbf {\bibinfo {volume}
  {53}},\ \bibinfo {pages} {1} (\bibinfo {year} {2010})}\BibitemShut {NoStop}%
\bibitem [{\citenamefont {Sac{\'e}p{\'e}}\ \emph {et~al.}(2011)\citenamefont
  {Sac{\'e}p{\'e}}, \citenamefont {Dubouchet}, \citenamefont {Chapelier},
  \citenamefont {Sanquer}, \citenamefont {Ovadia}, \citenamefont {Shahar},
  \citenamefont {Feigel'man},\ and\ \citenamefont {Ioffe}}]{Sacepe:2011jm}%
  \BibitemOpen
  \bibfield  {author} {\bibinfo {author} {\bibfnamefont {B.}~\bibnamefont
  {Sac{\'e}p{\'e}}}, \bibinfo {author} {\bibfnamefont {T.}~\bibnamefont
  {Dubouchet}}, \bibinfo {author} {\bibfnamefont {C.}~\bibnamefont
  {Chapelier}}, \bibinfo {author} {\bibfnamefont {M.}~\bibnamefont {Sanquer}},
  \bibinfo {author} {\bibfnamefont {M.}~\bibnamefont {Ovadia}}, \bibinfo
  {author} {\bibfnamefont {D.}~\bibnamefont {Shahar}}, \bibinfo {author}
  {\bibfnamefont {M.~V.}\ \bibnamefont {Feigel'man}}, \ and\ \bibinfo {author}
  {\bibfnamefont {L.~B.}\ \bibnamefont {Ioffe}},\ }\href {\doibase
  10.1038/nphys1892} {\bibfield  {journal} {\bibinfo  {journal} {Nat Phys}\
  }\textbf {\bibinfo {volume} {6}},\ \bibinfo {pages} {1} (\bibinfo {year}
  {2011})}\BibitemShut {NoStop}%
\bibitem [{\citenamefont {Driessen}\ \emph {et~al.}(2012)\citenamefont
  {Driessen}, \citenamefont {Coumou}, \citenamefont {Tromp}, \citenamefont
  {de~Visser},\ and\ \citenamefont {Klapwijk}}]{Driessen:2012gx}%
  \BibitemOpen
  \bibfield  {author} {\bibinfo {author} {\bibfnamefont {E.~F.~C.}\
  \bibnamefont {Driessen}}, \bibinfo {author} {\bibfnamefont {P.~C. J.~J.}\
  \bibnamefont {Coumou}}, \bibinfo {author} {\bibfnamefont {R.~R.}\
  \bibnamefont {Tromp}}, \bibinfo {author} {\bibfnamefont {P.~J.}\ \bibnamefont
  {de~Visser}}, \ and\ \bibinfo {author} {\bibfnamefont {T.~M.}\ \bibnamefont
  {Klapwijk}},\ }\href {\doibase 10.1103/PhysRevLett.109.107003} {\bibfield
  {journal} {\bibinfo  {journal} {Phys. Rev. Lett.}\ }\textbf {\bibinfo
  {volume} {109}},\ \bibinfo {pages} {107003} (\bibinfo {year}
  {2012})}\BibitemShut {NoStop}%
\bibitem [{\citenamefont {Feigel'man}\ and\ \citenamefont
  {Skvortsov}(2012)}]{Feigelman:2012fp}%
  \BibitemOpen
  \bibfield  {author} {\bibinfo {author} {\bibfnamefont {M.~V.}\ \bibnamefont
  {Feigel'man}}\ and\ \bibinfo {author} {\bibfnamefont {M.~A.}\ \bibnamefont
  {Skvortsov}},\ }\href {\doibase 10.1103/PhysRevLett.109.147002} {\bibfield
  {journal} {\bibinfo  {journal} {Phys. Rev. Lett.}\ }\textbf {\bibinfo
  {volume} {109}},\ \bibinfo {pages} {147002} (\bibinfo {year}
  {2012})}\BibitemShut {NoStop}%
\bibitem [{\citenamefont {Diener}\ \emph {et~al.}(2012)\citenamefont {Diener},
  \citenamefont {LeDuc}, \citenamefont {Yates}, \citenamefont {Lankwarden},\
  and\ \citenamefont {Baselmans}}]{Diener:2012hk}%
  \BibitemOpen
  \bibfield  {author} {\bibinfo {author} {\bibfnamefont {P.}~\bibnamefont
  {Diener}}, \bibinfo {author} {\bibfnamefont {H.~G.}\ \bibnamefont {LeDuc}},
  \bibinfo {author} {\bibfnamefont {S.~J.~C.}\ \bibnamefont {Yates}}, \bibinfo
  {author} {\bibfnamefont {Y.~J.~Y.}\ \bibnamefont {Lankwarden}}, \ and\
  \bibinfo {author} {\bibfnamefont {J.~J.~A.}\ \bibnamefont {Baselmans}},\
  }\href {\doibase 10.1007/s10909-012-0484-z} {\bibfield  {journal} {\bibinfo
  {journal} {J. Low Temp. Phys.}\ }\textbf {\bibinfo {volume} {167}},\ \bibinfo
  {pages} {305} (\bibinfo {year} {2012})}\BibitemShut {NoStop}%
\bibitem [{\citenamefont {Iossad}\ \emph {et~al.}(2000)\citenamefont {Iossad},
  \citenamefont {Mijiritskii}, \citenamefont {Roddatis}, \citenamefont {van~der
  Pers}, \citenamefont {Jackson}, \citenamefont {Gao}, \citenamefont
  {Polyakov}, \citenamefont {Dmitriev},\ and\ \citenamefont
  {Klapwijk}}]{Iosad:2000gz}%
  \BibitemOpen
  \bibfield  {author} {\bibinfo {author} {\bibfnamefont {N.~N.}\ \bibnamefont
  {Iossad}}, \bibinfo {author} {\bibfnamefont {A.~V.}\ \bibnamefont
  {Mijiritskii}}, \bibinfo {author} {\bibfnamefont {V.~V.}\ \bibnamefont
  {Roddatis}}, \bibinfo {author} {\bibfnamefont {N.~M.}\ \bibnamefont {van~der
  Pers}}, \bibinfo {author} {\bibfnamefont {B.~D.}\ \bibnamefont {Jackson}},
  \bibinfo {author} {\bibfnamefont {J.~R.}\ \bibnamefont {Gao}}, \bibinfo
  {author} {\bibfnamefont {S.~N.}\ \bibnamefont {Polyakov}}, \bibinfo {author}
  {\bibfnamefont {P.~N.}\ \bibnamefont {Dmitriev}}, \ and\ \bibinfo {author}
  {\bibfnamefont {T.~M.}\ \bibnamefont {Klapwijk}},\ }\href {\doibase
  10.1063/1.1319653} {\bibfield  {journal} {\bibinfo  {journal} {J. Appl.
  Phys.}\ }\textbf {\bibinfo {volume} {88}},\ \bibinfo {pages} {5756} (\bibinfo
  {year} {2000})}\BibitemShut {NoStop}%
\bibitem [{Note1()}]{Note1}%
  \BibitemOpen
  \bibinfo {note} {In the case of film C, the fabrication process was slightly
  different, resulting in the TiN layer still being present beneath the NbTiN
  layer.}\BibitemShut {Stop}%
\bibitem [{\citenamefont {Coumou}\ \emph {et~al.}(2013)\citenamefont {Coumou},
  \citenamefont {Zuiddam}, \citenamefont {Driessen}, \citenamefont {de~Visser},
  \citenamefont {Baselmans},\ and\ \citenamefont {Klapwijk}}]{Coumou:gg}%
  \BibitemOpen
  \bibfield  {author} {\bibinfo {author} {\bibfnamefont {P.~C. J.~J.}\
  \bibnamefont {Coumou}}, \bibinfo {author} {\bibfnamefont {M.~R.}\
  \bibnamefont {Zuiddam}}, \bibinfo {author} {\bibfnamefont {E.~F.~C.}\
  \bibnamefont {Driessen}}, \bibinfo {author} {\bibfnamefont {P.~J.}\
  \bibnamefont {de~Visser}}, \bibinfo {author} {\bibfnamefont {J.~J.~A.}\
  \bibnamefont {Baselmans}}, \ and\ \bibinfo {author} {\bibfnamefont {T.~M.}\
  \bibnamefont {Klapwijk}},\ }\href {\doibase 10.1109/TASC.2012.2236603}
  {\bibfield  {journal} {\bibinfo  {journal} {Ieee T Appl Supercon}\ }\textbf
  {\bibinfo {volume} {23}},\ \bibinfo {pages} {7500404} (\bibinfo {year}
  {2013})}\BibitemShut {NoStop}%
\bibitem [{\citenamefont {Zmuidzinas}(2011)}]{Zmuidzinas:2012wj}%
  \BibitemOpen
  \bibfield  {author} {\bibinfo {author} {\bibfnamefont {J.}~\bibnamefont
  {Zmuidzinas}},\ }\href {https://webmail.cea.fr/owa/} {\bibfield  {journal}
  {\bibinfo  {journal} {Annu. Rev. Cond. Mat. Phys.}\ }\textbf {\bibinfo
  {volume} {3}},\ \bibinfo {pages} {169} (\bibinfo {year} {2011})}\BibitemShut
  {NoStop}%
\bibitem [{\citenamefont {supplementary information at [] for~detailed
  information.}()}]{supplementary}%
  \BibitemOpen
  \bibfield  {author} {\bibinfo {author} {\bibfnamefont {See}~\bibnamefont
  {supplementary information at ... for detailed information}}}\href@noop {}
  {\ }\BibitemShut {NoStop}%
\bibitem [{Note2()}]{Note2}%
  \BibitemOpen
  \bibinfo {note} {This film is labelled $E$ in Ref.\protect \rev@citealpnum
  {Driessen:2012gx}}\BibitemShut {NoStop}%
\bibitem [{\citenamefont {Baturina}\ \emph {et~al.}(2008)\citenamefont
  {Baturina}, \citenamefont {Bilu{\v s}i{\'c}}, \citenamefont {Mironov},
  \citenamefont {Vinokur}, \citenamefont {Baklanov},\ and\ \citenamefont
  {Strunk}}]{Baturina:2008ww}%
  \BibitemOpen
  \bibfield  {author} {\bibinfo {author} {\bibfnamefont {T.~I.}\ \bibnamefont
  {Baturina}}, \bibinfo {author} {\bibfnamefont {A.}~\bibnamefont {Bilu{\v
  s}i{\'c}}}, \bibinfo {author} {\bibfnamefont {A.~Y.}\ \bibnamefont
  {Mironov}}, \bibinfo {author} {\bibfnamefont {V.~M.}\ \bibnamefont
  {Vinokur}}, \bibinfo {author} {\bibfnamefont {M.~R.}\ \bibnamefont
  {Baklanov}}, \ and\ \bibinfo {author} {\bibfnamefont {C.}~\bibnamefont
  {Strunk}},\ }\href
  {http://www.sciencedirect.com/science/article/pii/S0921453407013615}
  {\bibfield  {journal} {\bibinfo  {journal} {Physica C: Superconductivity}\
  }\textbf {\bibinfo {volume} {468}},\ \bibinfo {pages} {316} (\bibinfo {year}
  {2008})}\BibitemShut {NoStop}%
\bibitem [{\citenamefont {Khalil}\ \emph {et~al.}(2012)\citenamefont {Khalil},
  \citenamefont {Stoutimore}, \citenamefont {Wellstood},\ and\ \citenamefont
  {Osborn}}]{Khalil:2012jr}%
  \BibitemOpen
  \bibfield  {author} {\bibinfo {author} {\bibfnamefont {M.~S.}\ \bibnamefont
  {Khalil}}, \bibinfo {author} {\bibfnamefont {M.~J.~A.}\ \bibnamefont
  {Stoutimore}}, \bibinfo {author} {\bibfnamefont {F.~C.}\ \bibnamefont
  {Wellstood}}, \ and\ \bibinfo {author} {\bibfnamefont {K.~D.}\ \bibnamefont
  {Osborn}},\ }\href {\doibase 10.1063/1.3692073} {\bibfield  {journal}
  {\bibinfo  {journal} {J. Appl. Phys.}\ }\textbf {\bibinfo {volume} {111}},\
  \bibinfo{pages}{054510} (\bibinfo {year} {2012})}\BibitemShut {NoStop}%
\bibitem [{\citenamefont {Sac{\'e}p{\'e}}\ \emph {et~al.}(2008)\citenamefont
  {Sac{\'e}p{\'e}}, \citenamefont {Chapelier}, \citenamefont {Baturina},
  \citenamefont {Vinokur}, \citenamefont {Baklanov},\ and\ \citenamefont
  {Sanquer}}]{Sacepe:2008jx}%
  \BibitemOpen
  \bibfield  {author} {\bibinfo {author} {\bibfnamefont {B.}~\bibnamefont
  {Sac{\'e}p{\'e}}}, \bibinfo {author} {\bibfnamefont {C.}~\bibnamefont
  {Chapelier}}, \bibinfo {author} {\bibfnamefont {T.~I.}\ \bibnamefont
  {Baturina}}, \bibinfo {author} {\bibfnamefont {V.~M.}\ \bibnamefont
  {Vinokur}}, \bibinfo {author} {\bibfnamefont {M.~R.}\ \bibnamefont
  {Baklanov}}, \ and\ \bibinfo {author} {\bibfnamefont {M.}~\bibnamefont
  {Sanquer}},\ }\href {\doibase 10.1103/PhysRevLett.101.157006} {\bibfield
  {journal} {\bibinfo  {journal} {Phys. Rev. Lett.}\ }\textbf {\bibinfo
  {volume} {101}},\  \bibinfo{pages}{157006} (\bibinfo {year} {2008})}\BibitemShut {NoStop}%
\bibitem [{\citenamefont {Nam}(1967)}]{Nam:1967ue}%
  \BibitemOpen
  \bibfield  {author} {\bibinfo {author} {\bibfnamefont {S.}~\bibnamefont
  {Nam}},\ }\href
  {http://links.isiglobalnet2.com/gateway/Gateway.cgi?GWVersion=2&SrcAuth=mekentosj&SrcApp=Papers&DestLinkType=FullRecord&DestApp=WOS&KeyUT=A19679348900030}
  {\bibfield  {journal} {\bibinfo  {journal} {Phys. Rev.}\ }\textbf {\bibinfo
  {volume} {156}},\ \bibinfo {pages} {470} (\bibinfo {year}
  {1967})}\BibitemShut {NoStop}%
\bibitem [{\citenamefont {Altshuler}\ and\ \citenamefont
  {Aronov}(1985)}]{Altshuler:1985tw}%
  \BibitemOpen
  \bibfield  {author} {\bibinfo {author} {\bibfnamefont {B.~L.}\ \bibnamefont
  {Altshuler}}\ and\ \bibinfo {author} {\bibfnamefont {A.~G.}\ \bibnamefont
  {Aronov}},\ }\href
  {http://lib.org.by/info/P_Physics/PS_Solid%20state/PSa_Applications/Altshuler%20B.L.,%20Aronov%20A.G.%20Electron-electron%20interation%20in%20disordered%20conductors%20(Elsevier,%201985)(400dpi)(T)(153s).djvu}
  {\emph {\bibinfo {title} {{Electron-electron Interaction in Disordered
  Conductors}}}}\ (\bibinfo  {publisher} {Elsevier},\ \bibinfo {year}
  {1985})\BibitemShut {NoStop}%
\bibitem [{\citenamefont {Pierre}\ \emph {et~al.}(2001)\citenamefont {Pierre},
  \citenamefont {Pothier}, \citenamefont {Joyez}, \citenamefont {Birge},
  \citenamefont {Esteve},\ and\ \citenamefont {Devoret}}]{Pierre:2001fo}%
  \BibitemOpen
  \bibfield  {author} {\bibinfo {author} {\bibfnamefont {F.}~\bibnamefont
  {Pierre}}, \bibinfo {author} {\bibfnamefont {H.}~\bibnamefont {Pothier}},
  \bibinfo {author} {\bibfnamefont {P.}~\bibnamefont {Joyez}}, \bibinfo
  {author} {\bibfnamefont {N.~O.}~\bibnamefont {Birge}}, \bibinfo {author}
  {\bibfnamefont {D.}~\bibnamefont {Esteve}}, \ and\ \bibinfo {author}
  {\bibfnamefont {M.~H.}\ \bibnamefont {Devoret}},\ }\href {\doibase
  10.1103/PhysRevLett.86.1590} {\bibfield  {journal} {\bibinfo  {journal}
  {Phys. Rev. Lett.}\ }\textbf {\bibinfo {volume} {86}},\ \bibinfo {pages}
  {1590} (\bibinfo {year} {2001})}\BibitemShut {NoStop}%
\bibitem [{\citenamefont {Lemari{\'e}}\ \emph {et~al.}(2013)\citenamefont
  {Lemari{\'e}}, \citenamefont {Kamlapure}, \citenamefont {Bucheli},
  \citenamefont {Benfatto}, \citenamefont {Lorenzana}, \citenamefont {Seibold},
  \citenamefont {Ganguli}, \citenamefont {Raychaudhuri},\ and\ \citenamefont
  {Castellani}}]{Lemarie:2013bk}%
  \BibitemOpen
  \bibfield  {author} {\bibinfo {author} {\bibfnamefont {G.}~\bibnamefont
  {Lemari{\'e}}}, \bibinfo {author} {\bibfnamefont {A.}~\bibnamefont
  {Kamlapure}}, \bibinfo {author} {\bibfnamefont {D.}~\bibnamefont {Bucheli}},
  \bibinfo {author} {\bibfnamefont {L.}~\bibnamefont {Benfatto}}, \bibinfo
  {author} {\bibfnamefont {J.}~\bibnamefont {Lorenzana}}, \bibinfo {author}
  {\bibfnamefont {G.}~\bibnamefont {Seibold}}, \bibinfo {author} {\bibfnamefont
  {S.~C.}\ \bibnamefont {Ganguli}}, \bibinfo {author} {\bibfnamefont
  {P.}~\bibnamefont {Raychaudhuri}}, \ and\ \bibinfo {author} {\bibfnamefont
  {C.}~\bibnamefont {Castellani}},\ }\href {\doibase
  10.1103/PhysRevB.87.184509} {\bibfield  {journal} {\bibinfo  {journal} {Phys
  Rev B}\ }\textbf {\bibinfo {volume} {87}},\ \bibinfo {pages} {184509}
  (\bibinfo {year} {2013})}\BibitemShut {NoStop}%
\bibitem [{\citenamefont {Ghosal}\ \emph {et~al.}(1998)\citenamefont {Ghosal},
  \citenamefont {Randeria},\ and\ \citenamefont {Trivedi}}]{Ghosal:1998up}%
  \BibitemOpen
  \bibfield  {author} {\bibinfo {author} {\bibfnamefont {A.}~\bibnamefont
  {Ghosal}}, \bibinfo {author} {\bibfnamefont {M.}~\bibnamefont {Randeria}}, \
  and\ \bibinfo {author} {\bibfnamefont {N.}~\bibnamefont {Trivedi}},\ }\href
  {http://links.isiglobalnet2.com/gateway/Gateway.cgi?GWVersion=2&SrcAuth=mekentosj&SrcApp=Papers&DestLinkType=FullRecord&DestApp=WOS&KeyUT=000076829500034}
  {\bibfield  {journal} {\bibinfo  {journal} {Phys. Rev. Lett.}\ }\textbf
  {\bibinfo {volume} {81}},\ \bibinfo {pages} {3940} (\bibinfo {year}
  {1998})}\BibitemShut {NoStop}%
\bibitem [{\citenamefont {Ghosal}\ \emph {et~al.}(2002)\citenamefont {Ghosal},
  \citenamefont {Randeria},\ and\ \citenamefont {Trivedi}}]{Ghosal:2002cn}%
  \BibitemOpen
  \bibfield  {author} {\bibinfo {author} {\bibfnamefont {A.}~\bibnamefont
  {Ghosal}}, \bibinfo {author} {\bibfnamefont {M.}~\bibnamefont {Randeria}}, \
  and\ \bibinfo {author} {\bibfnamefont {N.}~\bibnamefont {Trivedi}},\ }\href
  {\doibase 10.1103/PhysRevB.65.014501} {\bibfield  {journal} {\bibinfo
  {journal} {Phys. Rev. B}\ }\textbf {\bibinfo {volume} {65}},\  \bibinfo {pages} {014501} (\bibinfo {year}
  {2001})}\BibitemShut {NoStop}%
\bibitem [{\citenamefont {Sherman}\ \emph {et~al.}(2013)\citenamefont
  {Sherman}, \citenamefont {Gorshunov}, \citenamefont {Poran}, \citenamefont
  {Jesudasan}, \citenamefont {Raychaudhuri}, \citenamefont {Trivedi},
  \citenamefont {Dressel},\ and\ \citenamefont {Frydman}}]{Sherman:2013ta}%
  \BibitemOpen
  \bibfield  {author} {\bibinfo {author} {\bibfnamefont {D.}~\bibnamefont
  {Sherman}}, \bibinfo {author} {\bibfnamefont {B.}~\bibnamefont {Gorshunov}},
  \bibinfo {author} {\bibfnamefont {S.}~\bibnamefont {Poran}}, \bibinfo
  {author} {\bibfnamefont {J.}~\bibnamefont {Jesudasan}}, \bibinfo {author}
  {\bibfnamefont {P.}~\bibnamefont {Raychaudhuri}}, \bibinfo {author}
  {\bibfnamefont {N.}~\bibnamefont {Trivedi}}, \bibinfo {author} {\bibfnamefont
  {M.}~\bibnamefont {Dressel}}, \ and\ \bibinfo {author} {\bibfnamefont
  {A.}~\bibnamefont {Frydman}},\ }\href {http://arxiv.org/abs/1304.7087v1}
  {\bibfield  {journal} {\bibinfo  {journal} {arXiv}\ } (\bibinfo {year}
  {2013})},\ \Eprint {http://arxiv.org/abs/1304.7087v1} {1304.7087v1}
  \BibitemShut {NoStop}%
\bibitem [{\citenamefont {Gu{\'e}ron}\ \emph {et~al.}(1996)\citenamefont
  {Gu{\'e}ron}, \citenamefont {Pothier}, \citenamefont {Birge}, \citenamefont
  {Esteve},\ and\ \citenamefont {Devoret}}]{Gueron:1996hw}%
  \BibitemOpen
  \bibfield  {author} {\bibinfo {author} {\bibfnamefont {S.}~\bibnamefont
  {Gu{\'e}ron}}, \bibinfo {author} {\bibfnamefont {H.}~\bibnamefont {Pothier}},
  \bibinfo {author} {\bibfnamefont {N.~O.}~\bibnamefont {Birge}}, \bibinfo
  {author} {\bibfnamefont {D.}~\bibnamefont {Esteve}}, \ and\ \bibinfo {author}
  {\bibfnamefont {M.~H.}\ \bibnamefont {Devoret}},\ }\href {\doibase
  10.1103/PhysRevLett.77.3025} {\bibfield  {journal} {\bibinfo  {journal}
  {Phys. Rev. Lett.}\ }\textbf {\bibinfo {volume} {77}},\ \bibinfo {pages}
  {3025} (\bibinfo {year} {1996})}\BibitemShut {NoStop}%
\end{thebibliography}
%

\end{document}